\documentclass[twocolumn,showpacs,prl,amsmath,floatfix]{revtex4}
\usepackage{graphicx}
\begin{document}

\title{Theory of Insulator Metal Transition and Colossal Magnetoresistance in Doped Manganites}

\author{T.V. Ramakrishnan\cite{add1}, H. R. Krishnamurthy\cite{add1}, S.R.Hassan\cite{add1}
and G. V. Pai\cite{add2} }
\address{Centre for Condensed Matter Theory, Department of Physics,
Indian Institute of Science, Bangalore 560 012, India}
\date{\today}
\begin{abstract}
The persistent proximity of insulating and metallic phases, a
puzzling characterestic of manganites, is argued to arise from the
self organization of the twofold degenerate $e_g$ orbitals of $Mn$
into localized Jahn-Teller(JT) polaronic levels and broad band
states due to the large electron - JT phonon coupling present in
them. We describe a new two band model with strong correlations
and a dynamical mean-field theory calculation of equilibrium and
transport properties. These explain the insulator metal transition
and colossal magnetoresistance quantitatively, as well as other
consequences of two state coexistence.
\end{abstract}
\pacs{75.47.Lx, 75.47.Gk, 71.27.+a, 71.30.+h, 71.38.-k}
\maketitle

Doped perovskite manganites $Re_{1-x} A_x Mn O_3$, where $Re$ and
$A$ are rare earth and alkaline earth ions, show a rich variety of
electronic, magnetic and structural phenomena and phases
\cite{reviews,Dagotto}. Unusual effects, such as
insulator-metal(IM) transitions both as a function of $x$ and of
temperature $T$ over a wide region $x \stackrel {<}{\sim}$ 0.5, or
even as a consequence of isotope substitution ($O^{18}\rightarrow
O^{16}$)\cite{IMT-iso}, colossal magneto-resistance(CMR) near
$T_{IM}$ and 'melting' of the charge/orbitally ordered insulator
$(x \stackrel {>}{\sim} 0.5)$ into a metal in a relatively
small(5-7 $Tesla$) magnetic field all suggest that metallic and
insulating phases are always very close in free energy. This is
also reflected in the ubiquitous coexistence (static or dynamic)
of two 'phases', one insulating with local lattice distortion and
the other metallic without lattice distortion, with length scales
varying from $10 A^{\circ}$ to $10^3 A^{\circ}$ \cite{Dagotto}.

These phenomena are due to the dynamics of the $e_g$ electrons of
$Mn$ constrained by three strong on-site interactions, namely
electron lattice or Jahn-Teller (JT) coupling which splits the
twofold $e_g$ orbital degeneracy, ferromagnetic $e_g$
spin-$t_{2g}$ spin exchange or Hund's coupling $J_H$ and $e_g$
electron repulsion $U$. The respective energies are $E_{JT}\simeq
0.5 \, eV$, $J_H \simeq 2 \, eV$ and $U\simeq 5 \, eV$, compared
to the $e_g$ electron intersite hopping $t \simeq 0.2 \, eV$ which
sets the kinetic energy scale \cite{SandS}. Understanding their
observed consequences is one of the major challenges in the
physics of strongly interacting electrons. Earlier theoretical
attempts neglect one or more of these strong interactions and make
further approximations; the predictions do not agree with many
characteristics of manganites. For example the ferromagnetic Curie
transition in a number of manganites is from an insulator to a
metal, for $0.2 \stackrel {<}{\sim} x \stackrel {<}{\sim} 0.5$.
However, a theory with just Hund's coupling, due to Furukawa
\cite{Furukawa}, finds only a metallic phase, while Millis, Muller
and Shraiman \cite{MMS}, who additionally include the electron-JT
phonon coupling $g$ but treat the JT distortions as static
displacements, obtain a metal-metal transition crossing over to an
insulator-insulator transition as $g$ increases.

We propose and implement here a new approach which incorporates
the crucial effects of all the three interactions and is based on
a new idea, namely that of {\em{coexisting $JT$ polaronic and
broad band $e_g$ states}}, which we believe is the key to
manganite physics. The idea and some of its consequences are
described and calculations based on a {\em{new two band model}}
are then outlined.

We first discuss the  effect of large JT coupling $g$ on the
initially twofold degenerate $e_g$ orbitals at each lattice site.
There is one superposition (labelled $\ell$) which, when singly
occupied,leads to a polaronic state with local octahedral symmetry
breaking $Mn-O$ bond distortion and energy $-E_{JT}$. Its
intersite hopping is reduced (for $(E_{JT}/\hbar \omega_0)\gg 1$
where $\omega_0$ is the JT phonon mode frequency) by the
exponential Huang-Rhys \cite{HRfac} or phonon overlap factor $\eta
\simeq exp \left(-E_{JT}/2 \hbar \omega_0\right)$. Since $\eta
\simeq exp (-5)$ for $La Mn O_3 $, and being a local quantity is
not likely to change much on doping, the $\ell$ polaron bandwidth
$2D^{\ast} \sim 2z t \eta \simeq k_B (125K)$ is small, and is
neglected in much of this paper. At each site, there is
necessarily another, orthogonal state (labelled $b$). Its energy
is zero on the fraction $x$ of sites where the $\ell$ electrons
are not present(hole-sites), and is $\bar U = (U+2E_{JT} )$ on the
$\ell$ electron sites. The $b$ electron hops adiabatically
$\left(\hbar t^{-1} \ll \omega_0^{-1}\right)$, in an annealed
random medium of repulsive $\ell$ sites and (for a homogeneous
orbital liquid) forms a band whose re-normalized bandwidth $2D$
increases with $x$ as well as with $T^{-1}$ and $H$. For, the
inhibition of $b$ hopping due to large $\bar U$ is reduced when
there are more hole-sites, and that due to large $J_H$ is reduced
when the $t_{2g}$ spin order is enhanced.We show in this paper
that the simultaneous presence of $\ell$ polaron states and the
$b$ band, the change in the effective band width of the latter
with $x, T, H$ etc. and the consequent change in their relative
occupation (leading to an insulator eg for $E_{JT}
> D$ and metal for $E_{JT} < D$ ) are the basis of many phenomena
seen in manganites. Our picture requires the persistence of local
$JT$ distortions well into the metallic regime, as indeed seen by
direct probes of instantaneous $Mn-O$ bond length
\cite{pnd-exafs}. The $\ell$ polaron motion is anti-adiabatic
$(2D^{\ast} \ll \hbar \omega_0 )$ while the $b$ electron motion is
adiabatic $(2D \gg \hbar \omega_0 )$ ; the respective bandwidths
are exponentially small and of order the bare value. By contrast,
in earlier theoretical work \cite{MMS,craco} where the $JT$
distortion is treated as static, both $\ell$ and $b$ electron
dynamics become adiabatic so that the bands are of comparable
width, leading to consequences which disagree with experiment.

The system of correlated $\ell$ and $b$ states described above can
be modelled by the Hamiltonian
\begin{eqnarray}
H_{\ell b} =( -E_{JT}-\mu ) \sum_{i,\sigma} \ell_{i
\sigma}^\dagger \ell_{i \sigma} -\mu \sum_{i\sigma}
b_{i\sigma}^\dagger b_{i\sigma}
\\\nonumber - \bar {t}\sum_{\langle ij \rangle , \sigma} (b_{i
\sigma}^\dagger b_{j\sigma} + hc )+ \bar {U} \sum_{i\sigma}
n_{\ell i\sigma} n_{b i\sigma} + H_m
\end{eqnarray}
In Eq.(1), the $\ell$ polaron has energy $(-E_{JT})$, the $b$
electrons hop between nearest neighbour sites with an effective
amplitude $\bar t$, and $\bar U$ is the effective repulsion
between $\ell$ polarons and $b$ electrons of the same spin at a
particular site \cite{n-neglectU}. The common chemical potential
is $\mu$. $H_m$ is the magnetic part involving the $e_g$ spins
$\vec {s_i}$ and $t_{2g}$ core spins $\vec {S_i}$, namely
\begin{eqnarray}
H_m=-J_H \sum_i \vec{s_i} \cdot \vec {S_i} -J_F \sum_{\langle ij
\rangle } \vec {S_i} \cdot \vec {S_j} -\mu_B \sum_i \vec {S_i}
\cdot \vec H
\end{eqnarray}
In addition to the familiar  Hund's coupling $J_H$ and Zeeman
coupling to an external field $\vec {H}$ (neglecting the
relatively smaller term $\mu_B \vec {s_i}\cdot \vec {H}$) we have
included in $H_m$ a {\em{new ferromagnetic nearest neighbour
exchange $J_F$ between the $t_{2g}$ core spins}}.

The new term arises from a {\em{virtual double-exchange process}}
in which an $\ell$ electron at a site $i$ hops
quickly(adiabatically) to a nearest neighbour site $j$ and back.
In the limit of large $\bar U$ and $J_H$ , and approximating the
$t_{2g}$ spins as classical, i.e., $\vec{S}_i = S \hat{\Omega}_i$
where $\hat{\Omega}_i$ are unit vectors, the energy shift to
second order in $\bar t$ is
$\left(\bar {t}^2 / 2E_{JT} \right) \frac{1}{2}({\hat \Omega}_i
\cdot {\hat \Omega}_j + 1 )\left(n_{\ell i} (1-n_j) + n_{\ell j}
(1-n_i)\right)$.
The energy denominator $2E_{JT}$ arises from the unrelaxed JT
distorted intermediate state, the $\frac{1}{2}({\hat \Omega}_i
\cdot {\hat \Omega}_j + 1 )$ factor from large $J_H$, and the
occupancy dependent terms from large $\bar U$. We have further
approximated this by the simple exchange term in Eq.(2), whence
$J_F \simeq \left(\bar {t}^2 / ({2E_{JT} S^2}) \right) x(1-x)$.

The model Hamiltonian $H_{\ell b}$ (Eq.(1)) can be motivated
starting from a lattice model with two $e_g$ orbitals per
site\cite{reviews,Dagotto}. Let $a_{i\alpha}^{\dagger}$ create an
electron in orbital $\alpha$ at site $i$, ( with $|\alpha = 1 >
\equiv |x^2-y^2>$ and $ |\alpha = 2> \equiv |3z^2-r^2>$ say),
locally coupled to JT lattice modes $(Q_{xi}, Q_{zi}) \equiv (Q_i
, \theta_i )$ by the term $H_{JT}^i = g a_{i\alpha}^{\dagger}\vec
{\tau}_{\alpha\beta} a_{i\beta} \cdot \vec {Q}_i$ where $\vec
{\tau}$ are the Pauli matrices. The eigenvalues of $H_{JT}^i$ are
$\pm g Q_i$ , the corresponding electron creation operators  being
labelled $b^{\dagger}$  and ${\tilde {\ell}}^{\dagger}$. The
effective lattice potential energies for single electron
occupation of these states are $(K/2)Q_i^2 \pm g Q_i$ where $K$ is
the phonon force constant. The lower ($\ell$) branch has a minimum
at a static lattice distortion $Q_0= g/K$ and energy $-E_{JT} =
-(g^2/2K)$. $\ell^\dagger_i$ in Eq.(1) are polaron creation
operators given by $\hat{\eta_i}{\tilde{\ell}}^\dagger_i $ where
$\hat{\eta_i}$ ( $\sim exp(iQ_0 P_i / \hbar)$ where $P_i$ is the
radial momentum conjugate to $Q_i$ ) are unitary Lang-Firsov
\cite{L-F} like displacement transformations. The exponential
reduction $\eta$ in the hopping amplitude of $\ell$ arises
essentially from their ground state average. (We neglect the small
fluctuations with respect to this average). No lattice distortion
or reduction in hopping is associated with the higher ($b$)
branch. The small leading intersite term $H'_{\ell b} \sim \bar
{t} \eta \sum_{\langle ij\rangle } \left(\ell_i^{\dagger} b_j +
b_i^{\dagger} \ell_j + hc \right)$ describing $\ell-b$
hybridization is not included in Eq.(1); some consequences are
discussed later. Thus for $(E_{JT}/\hbar \omega_0)\gg 1$, and
neglecting $H'_{lb}$, we are led to Eq (1). We have made further
approximations which are realistic for a {\em{homogenous orbital
liquid}}, by statistically averaging quantities which depend on
the orbital admixture angle $\{\theta_i \}$, eg the intersite
hopping and the nearest neighbour $Mn-O$ bond correlations
(short-long), so that $\bar{t}$ and $b_i^\dagger $ are to be
regarded as averaged over $\{ \theta_i\}$. This is reasonable for
$0.2 \stackrel {<}{\sim} x \stackrel {<}{\sim} 0.5$ in most
manganites, but poor for those other values of $x$ for which one
has strong orbital correlations or long range orbital order.
Finally, in this regime, the chemical potential $\mu$ is chosen
such that
\begin{eqnarray}
\langle n_{\ell i}\rangle +\langle n_{bi}\rangle = \bar{n}_\ell +
\bar {n}_b = (1-x)
\end{eqnarray}

We have calculated the equilibrium and transport properties of
$H_{\ell b}$ (Eq.(1)) using the dynamical mean field theory (DMFT)
\cite{DMFT-rev} which is exact in $d=\infty$, and is accurate in
three dimensions if spatial correlations are not crucial. $H_{\ell
b}$, being a generalized form of the Falicov-Kimball
model\cite{FKM} with additional spin interactions $H_m$ (Eq.(2))
and constraint Eq.(3), is exactly soluble in this limit
\cite{DMFT-rev}.  We treat the $t_{2g}$ core spins as classical as
stated earlier, and their interaction in the Curie-Weiss mean
field approximation, and work in the large $J_H$ limit (i.e.
$SJ_H/\bar{t}\gg 1$). The local self energy $\Sigma_{ii}
(\omega^+)$ of the $b$ electrons and the mean magnetization $m=
\langle \hat{\Omega}_i \rangle$ are determined self-consistently.
From these we calculate the spectral density or density of states
(DOS) $\rho_b(\omega)$ of the $b$ electrons which determines their
occupation relative to that of the $\ell$ state, the
current-current correlation function relevant for the Kubo formula
for the electrical conductivity, etc\cite{n-details}. Most of our
calculations use a semicircular bare DOS with a bare bandwidth
$2D_o$\cite{n-dos}.

Our results for $\rho_b(\omega)$ for different values of $x$ and
$T$ are shown in Fig.1.  Since the effective bandwidth $2D$ of the
$b$ electrons decreases significantly as $x$ decreases (for any
sizeable $\bar{U}$), the $b$ band bottom is above the $\ell$ level
for small $x$ e.g $x=0.1$ (Fig.1a), and the low temperature state
is an insulator, ferromagnetic because of $J_F$. (We ignore the
anti-ferromagnetic super-exchange as it is dominated by the much
larger $J_F$ for $x \stackrel{>}\sim .1$ ). On increasing $x$, $D$
increases as well, and beyond a critical $x_c$ for which $D
=E_{JT}$, the low temperature state is a ferromagnetic metal, as
in Fig.1b. We note that while both $b$ and $\ell$ states are
occupied in the metal, most electrons are in the latter, polaronic
state since $\bar{n}_\ell = 0.66$ and $\bar{n}_b = 0.04$. $T_c$
{\em{is largely due to}} $J_{F}$; for, ${\bar n}_{b}$ being very
small, so is the $b$ electron contribution to $T_c$ via
conventional double exchange. On increasing $T$ at this $x$, the
$t_{2g}$ spins disorder, reducing the effective $b$ electron
hopping or $D$. Fig.1c shows the DOS at $T=350 K (> T_c)$.  It is
that of a small gap semiconductor. The metal insulator transition
occurs very near $T_c$ because of the strong feedback between $D$
and the magnetization $m$.  The carriers in the paramagnetic state
are $b$ electrons excited across the relatively small effective
band gap, not thermally unbound small polarons. The mobile carrier
fraction ${\bar n}_{b}$ is very small in the metallic phase
($\bar{n}_{b}\simeq 0.06$ for parameters corresponding to
$La_{0.7}Sr_{0.3}MnO_{3}$) and decreases rapidly\cite{n-details}
to a minimum at $T_c$, whereas the electron density is 0.7 per
site . This is exactly the hitherto unexplained inference for
$n_{eff}$ from the observed Drude weight in this compound
\cite{Neff}.

\begin{figure}
\includegraphics [height=5cm,width=8.5cm]{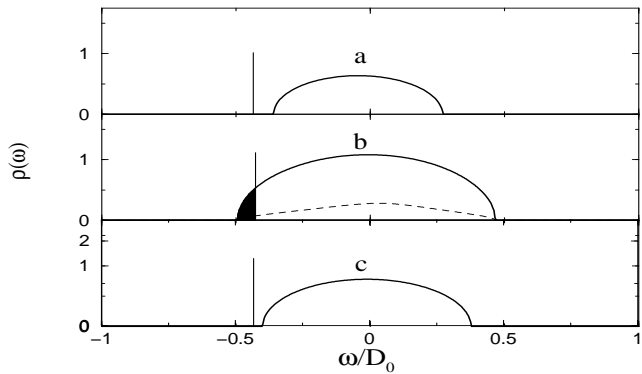}
\caption{Spectral density or $b$-DOS $\rho_b(\omega)$ for various
values of doping $x$ and temperature $T$. The effective $\ell$
polaron level is marked as a vertical line. Parameters chosen are
$E_{JT} = -0.5 \, eV$, $D_o=1.2 \, eV$ , $\bar U = 5.0 \, eV$ $J_F
= 2.23 \, meV $. (a) $x=0.1$, $T=0$, $\mu = -E_{JT}$;
ferromagnetic insulator. (b) $x=0.3$, $T=180 K (< T_c = 240K)$;
ferromagnetic metal. (c) $x=0.3$, $T=350 K$ ; paramagnetic
insulator. Full(dotted) lines correspond to up(down) spin DOS.
Occupied band states are shown shaded.}
\end{figure}

Our theory also describes transport properties fairly well. Fig.2
shows the electrical resistivity $\rho(T)$ versus $T$ for model
parameters chosen to fit $T_c$ and $\rho (T_c)$ for $La_{0.67}
Ca_{0.33} MnO_3$ \cite{LCMO-res}. The results for semicircular DOS
and tight binding DOS are nearly the same.  We see clearly the
sharp paramagnetic insulator to ferromagnetic metal transition,
the former having a calculated effective electrical gap
$\Delta_{eff}$ of 34 $meV$ while the experimental value is 48
$meV$ \cite{n-srcorr}. The resistivity falls dramatically to about
2 $m\Omega cm$ just below $T_c$ and does not decrease much further
thereafter, in contrast to the observed residual resistivity
values which can be as small as 50 $\mu\Omega cm$ . This is a
consequence of our neglect of inter-site $\ell$ coherence and is
discussed later. The resistivity in a field of 7 $Tesla$ is also
shown in Fig.2. The colossal decrease(CMR) is
apparent\cite{n-srcorr}.

\begin{figure}
\includegraphics [height=6cm,width=8.5cm]{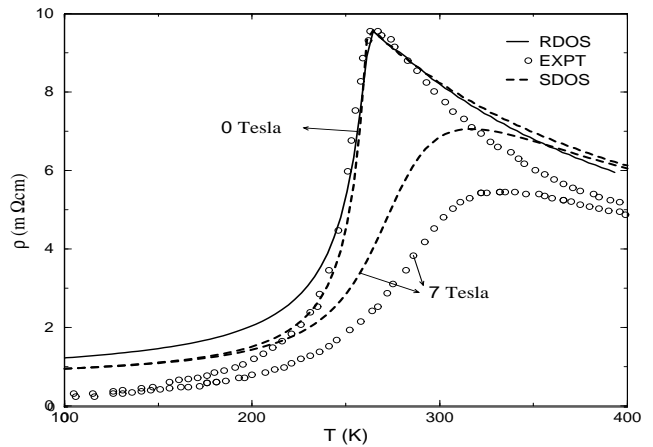}
\caption{The resistivity $\rho(T)$ of $La_{1-x}Ca_{x}MnO_{3}$
($x=0.33$) as a function of temperature $T$. Calculated (full
lines for realistic anisotropic hopping, dotted lines for a
semicircular DOS) and experimental (circles, from
ref.\cite{LCMO-res}) results for $H = 0$ and $7 \, Tesla$ are
shown. Parameters, chosen to fit $T_{c}$ and $\rho(T_{c})$, are
$E_{JT}= -0.5 \, eV$(SDOS),$ -0.6 \, eV$(RDOS), $2D_{o}= 2.4 \,
eV$ (SDOS), $2.44 \, eV$ (RDOS), $\bar U = 5 \, eV$, and $J_{F}=
2.37 \, meV$. } \label{Fig.2}
\end{figure}

The properties of manganites vary strongly and characteristically
with the ionic species $Re$ and $A$. This is natural in our model,
since the relative balance between metal and insulator is strongly
affected by small changes in $E_{JT}$ and $D_o$ and the carrier
density depends exponentially on $(D-E_{JT})$.  For example,
keeping $E_{JT}$ fixed (for a given $x=0.3$ say),and varying
$2D_o$, we find the variations in physical properties shown in
Fig.3. $T_c$, being proportional to $J_F$, increases roughly as
$D^2_o$ . The Curie transition changes from insulator-insulator to
insulator-metal to metal-metal, as is seen experimentally in the
sequence $Pr_{0.7}Ca_{0.3}MnO_3\,,\,La_{0.7}Ca_{0.3}MnO_3$ and
$La_{0.7}Sr_{0.3}MnO_3$ . $D_o$ is believed \cite{Dvsr-hwang} to
increase in this order. We also exhibit the enormous variation in
the calculated fractional magneto-resistance $\left\{\Delta
R(H)/R(H)\right\}_{T=T_c}$ for $H=7 \, Tesla$ as a function of
$T_c$; the agreement with experiment \cite{rel-cmr} is very good.

\begin{figure}
\includegraphics [height=7cm,width=8.5cm]{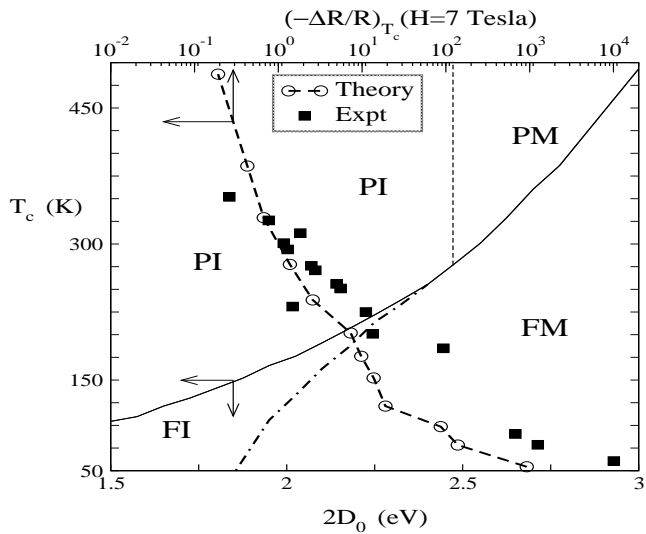}
\caption{Physical properties as a function of the $d$ electron
bandwidth $2D_{o}$, for fixed $E_{JT}= 0.5 \, eV$. Full line:
ferromagnetic $T_c$ vs $D_{o}$. The resistive transition is from
an insulator to an insulator(PI-FI) for small $D_{o}$, from an
insulator to a metal(PI-FM) for intermediate $D_{o}$, and from a
metal to a metal(PM-FM) for large $D_{o}$, as indicated . Broken
line: calculated fractional magnetoresistance ${(\Delta
R(H)/R(H))}_{T_c}$ vs $T_c$ for $H = 7 \, Tesla$ . Experimental
points are from \cite{rel-cmr}. }
\end{figure}

Thus, in the regime $0.2 < x <0.4$, a wide range of physical
properties for a number of doped manganites can be understood
physically and described quantitatively by our theory
\cite{n-details}.  The ratio $(E_{JT}/D_o)$ broadly determines the
systematics and $D_o$ the basic energy scale ($J_H$ and $\bar{U}$
being close to $\infty$ for our purposes).

The intersite hopping of the $\ell$ polarons arising from  term
$H'_{\ell b}$, with a characteristic temperature scale of $T^*
\sim (z\eta \bar{t}/k_B)\simeq 125K$, has many important
consequences. Below $T^*$ the $\ell$ states can form a coherent
band due to hybridization with $b$ states, whence the JT
distortion can become dynamic (time scale $\sim \hbar / k_BT^*$),
and also smaller self-consistently. The $b$ electron scattering
and the consequent resistivity then vanish at $T=0$, and are
nonzero only if static disorder is present. This can lead to a
metallic state with a small residual resistivity or to an Anderson
localized insulating state depending on the amount of disorder.
The strong dependence of $T_{c}$ on isotope mass $M_{0}$ is also a
consequence of $\ell$ polaron hopping, since its direct double
exchange contribution to $T_c$ will be proportional to ${\bar t}
(1-x) exp{( - E_{JT}/2\hbar\omega_{0})}$ and $\omega_{0}=\sqrt{K
/M_{0}}$. The isotope effect thus estimated is of the right size
\cite{zhao}.

Spatial correlations and inhomogeneities can be investigated in an
extended version of the model in Eq.(1) by including the orbital
angles $\theta_{i}$ as additional degrees of freedom. The
intersite hopping $t_{ij}$ now depends on $\theta_{i}$ and
$\theta_{j}$. There are a number of anharmonic, steric, elastic
terms, eg., nearest neighbour long-short $Mn-O$ bond correlations,
coupling between JT modes and strain,etc., which depend on these
angles as well. These, and other factors such as the variation of
$E_{JT}$ with local ion size, can be included in Eq.(1), and
questions such as short range correlation, orbital order, long
range order in $({\bar n}_{\ell i} \pm {\bar n}_{bi})$ (total
charge/relative charge order), disorder induced 'phase'
separation, explored by going beyond the self-consistent single
site DMFT used here. It would also be interesting to explore the
idea of exponentially separated time scales and adiabatic -
nonadiabatic crossover in  other oxides and organic molecules with
orbital degeneracy and strong local symmetry breaking Jahn Teller
coupling.

We would like to acknowledge support from the Indo-French Centre
for Promoting Advanced Research grant 2404-1 (HRK),US-India
project ONR N 00014-97-0988(TVR) and the Council for Scientific
and Industrial Research, India (SRH,GVP).


\begin{thebibliography}{99}
\bibitem[*]{add1} Also Condensed Matter Theory Unit, JNCASR, Jakkur,
Bangalore 560 064, India.

\bibitem[$\dagger$]{add2}Also Abdus Salam International Centre for Theoretical
Physics, 11 Strada Costiera, Trieste 34014, Italy

\bibitem{reviews} For reviews, see {\it Colossal Magnetoresistance Oxides}, ed Y
Tokura (Gordon and Breach, New York, 2000); M B Salamon and M Jaime, Rev. Mod.
Phys. {\bf 73}, 583 (2001).

\bibitem{Dagotto} eg. {\it Nanoscale Phase Separation in Manganites} by E
Dagotto (Springer Verlag, New York and Heidelberg, 2002).

\bibitem{IMT-iso} N A Babushkina,  {\it et al.}, Nature {\bf 391}, 159
(1998).

\bibitem{SandS} see eg. D D Sarma {\it et al.}, Phys. Rev. Lett.
{\bf 75}, 1126 (1995), S Satpathy, Z S Popovic, and F R
Vukajlovic, Phys. Rev. Lett. {\bf 76}, 960 (1996).

\bibitem{Furukawa} N Furukawa, Journal of the Physical
Society of Japan, {\bf 64} 2734 (1995).

\bibitem{MMS} A J Millis, R Mueller and B I Shraiman, Phys. Rev. B{\bf 54},
5389, 5405 (1995).

\bibitem{HRfac} K Huang and F Rhys,  Proc. Roy. Soc. London Ser. A {\bf 204}, 406
(1950).

\bibitem{pnd-exafs} see for example the pulsed neutron measurements of
D. Louca  {\it et al.},  Phys. Rev. B. {\bf 56}, R8475 (1997). and
the EXAFS results of C Meneghini {\it et al.}, J Phys. Cond. Matt.
{\bf 14}, 1967 (2002).

\bibitem{craco} M S Laad, L Craco and E Muller-Hartmann Phys. Rev. B {\bf
63}, 214419 (2001).

\bibitem{n-neglectU} We neglect same site opposite spin interactions since such
configurations are strongly disfavoured in the large $J_H$ limit
we treat in this paper.

\bibitem{L-F} G I Lang and A Yu Firsov, Soviet Physics JETP {\bf 43}, 1843 (1962).

\bibitem{DMFT-rev} A Georges, G Kotliar, W Krauth and M J Rozenberg, Rev. Mod. Phys.
{\bf 68}, 13 (1996).

\bibitem{FKM} L. M. Falicov and J C Kimball, Phys. Rev.
Lett. {\bf 22}, 997 (1969).

\bibitem{n-details} Details as well as
additional new results will be published elsewhere.

\bibitem{n-dos}We find that the calculated results
are nearly the same for semicircular DOS and realistic tight
binding DOS (eg. see Fig.2 for resistivity).

\bibitem{Neff} Y. Okimoto, et.al., Phys. Rev. B {\bf 55}, 4206
(1997).

\bibitem{LCMO-res} G. J. Snyder {\it et al.}, Phys. Rev. B {\bf 53}, 14434(1996).

\bibitem{n-srcorr}The discrepancies can be reduced
within our model by including short-range
correlations which are neglected in the DMFT.


\bibitem{Dvsr-hwang}H. Y. Hwang et. al., Phys. Rev.Lett.
{\bf 75}, 914 (1995).

\bibitem{rel-cmr} K Khazeni {\it et al.}, Phys. Rev. Lett. {\bf 76}, 295 (1996).

\bibitem{zhao} G. M. Zhao, {\it et al.}, Phys. Rev. B {\bf 63}, 060402 (2001).

\end{thebibliography}
\end{document}